\documentclass[a4paper,10pt]{article}
\usepackage[a4paper, margin=2cm]{geometry}

\usepackage{gensymb}
\usepackage{xcolor}
\usepackage{chemformula}
\usepackage[version=4]{mhchem}
\usepackage{subcaption}
\usepackage{graphicx} 
\usepackage{booktabs}

\usepackage{hyperref}  
\usepackage{natbib}
\usepackage{authblk}

\title{Before We Inject: Assessing the Impact of Silica-Based Aerosols on Stratospheric Chemistry via a Kinetic Model Informed by Molecular Dynamics}

 \author[1]{Dennis Lima}
 \author[1]{Saif Al-Kuwari}
 \affil[1]{Qatar Center for Quantum Computing, College of Science and Engineering, Hamad Bin Khalifa University Doha, Qatar}
 \author[2]{Ivan Gladich\footnote{ivan.gladich@uniurb.it}}
 \affil[2]{Department of Pure and Applied Sciences, 
Uniuversity of Urbino Carlo Bo, 61028, Urbino, Italy}

\date{July 2025}

\begin{document}

\maketitle

\begin{abstract}
Stratospheric aerosol injection (SAI) has been proposed as a geoengineering strategy to mitigate global warming by increasing Earth's albedo. Silica-based materials, such as diamond-doped silica aerogels, have shown promising optical properties, but their impact on stratospheric chemistry, ozone one in particular, remains largely unknown.  Here, we present \emph{first-principles} molecular dynamics (MD) simulations of the heterogeneous reaction between hydrogen chloride (\ch{HCl}) and chlorine nitrate (\ch{ClONO2}), two main reservoirs of stratospheric chlorine and nitrogen species, on a dry, hydroxylated $\alpha$-quartz silica interface, a surface that serves as a proxy for silica-based aerosols under low relative humidity and stratospheric conditions. Our results reveal a barrierless reaction pathway toward the formation of chlorine gas (\ch{Cl2}), a major contributor to stratospheric ozone loss. We design a heterogeneous kinetic model informed by our MD simulation and available experimental data: despite the barrierless formation of \ch{Cl2}, the higher surface affinities and partial pressures of \ch{HNO3} and \ch{HCl} compared to those of \ch{ClONO2} result in a negligible reaction probability, $\gamma_{\mathrm{ClONO_2}}$, upon chlorine nitrate collision with the silica surface. 

Since $\gamma_{\mathrm{ClONO_2}}$ enters as a proportionality constant in the definition of the heterogeneous reaction rate, our kinetic model indicates that the injection of silica-based aerosols may have only a limited impact on stratospheric ozone depletion driven by \ch{HCl} and \ch{ClONO2} chemistry.

 At the same time, our findings also underscore the scarcity of experimental data, the need of better theoretical frameworks for the inclusion of MD results into kinetic models, and the urgency for further experimental validations of silica-based SAI technologies before their deployment in climate intervention strategies.
\end{abstract}

\section{Introduction}

Anthropogenic greenhouse gas (GHG) emissions have been linked to global warming and climate change, posing significant environmental and societal challenges\cite{arias2021climate, jouzel2013brief}, and representing a serious threat to future generations. Even if emissions were immediately halted, the effects of global warming would persist for centuries. Although there is a massive ongoing research effort aimed at developing innovative solutions capable of removing GHGs directly from the atmosphere (e.g., direct air capture technologies\cite{custelcean2022direct, kumar2024adsorption, sinopoli2021stability, gladich2020tuning, martins2024structure}) or from industrial emissions,\cite{abotaleb2022chemical} current technologies remain inefficient and may not scale quickly enough to address the issue in a timely manner.\cite{clark2016consequences, national2015climate}

Stratospheric Aerosol Injection (SAI) is part of a broader category of solar geoengineering technologies and aims to reduce global warming by increasing Earth's albedo, i.e., by reflecting more solar radiation back into space and reducing the amount that reaches the surface. This concept was inspired by volcanic winter events observed after major eruptions, such as that of Mt. Pinatubo in 1991,\cite{bluth1992global, wilson1993situ} which emitted sulfur particles into the stratosphere and led to temporary global cooling. Inspired by this natural phenomenon, SAI has been proposed to help achieve the goal of the Paris Agreement of limiting global warming to $1.5\,\degree\mathrm{C}$.\cite{macmartin2018solar} However, despite its potential benefits, significant uncertainties remain regarding the risks associated with SAI, particularly its effects on stratospheric chemistry.\cite{robock200820,Keutsch2025}

Based on the observations of volcanic winters,  SAI research has initially focused on sulfur dioxide (\ch{SO2}) as the injected specie. However, \ch{SO2} oxidizes to form sulfuric acid aerosols, leading to optically inefficient aerosol size distributions \cite{vattioni2019exploring}, depletion of stratospheric ozone \cite{tilmes2008sensitivity, vattioni2019exploring, weisenstein2022interactive}, and stratospheric heating \cite{aquila2014modifications, dykema2016improved}. For these reasons, alternative substrates have begun to be explored\cite{dykema2016improved, ferraro2011stratospheric, keith2016stratospheric, pope2012stratospheric, weisenstein2015solar}  such as alumina (\ch{Al2O3}), calcite (\ch{CaCO3}),\cite{Keutsch2025} or silica-based materials doped with small diamonds.\cite{vukajlovic2021diamond} The overall objective is to identify materials that \emph{(a)} backscatter shortwave sunlight into the space, \emph{(b)} enable longwave radiation from Earth's surface to escape into space with minimal absorption (i.e., avoiding stratospheric heating), \emph{(c)} do not undergo rapid aging, and \emph{(d)} do not trigger undesired or harmful chemistry (e.g., stratospheric ozone loss). 

Among the undesired chemical processes that must be avoided, stratospheric ozone depletion is one of the most well-known, widely recognized also by the general public due to its association with the ozone hole episode at the end of the last century.\cite{farman1985large} Stratospheric ozone destruction is primarily driven by chlorine species, which can be activated by the following Reaction~\ref{r1} 
\begin{align} \label{r1}
\ch{HCl + ClONO2 <=> Cl2 + HNO3}
\end{align}
where hydrogen chloride (\ch{HCl}) and chlorine nitrate (\ch{ClONO2}) are important reservoirs of chlorine and nitrogen species in the stratosphere. Once \ch{Cl2} is formed, it can easily undergo photolysis under stratospheric conditions, leading to significant loss of ozone.

While Reaction~\ref{r1} has been extensively studied in the gas phase, its heterogeneous chemical pathways and the catalytic role (or lack thereof) of different substrates remain poorly understood. This knowledge gap represents one of the major concerns for the real-world implementation of SAI.
Molina et al.\cite{molina1997reaction} reported that Reaction~\ref{r1} can be catalyzed on the surface of alumina. Molina et al. conducted their experiments at stratospherically relevant temperatures, but they employed HCl concentrations that were an order of magnitude higher than those typically found in the stratosphere, likely because of the limited sensitivity of the experimental apparatus available at the time. Thus, extrapolating these findings to actual stratospheric conditions requires various assumptions, resulting
in large uncertainties (up to two orders of magnitude) in the heterogeneous reaction rate for Reaction~\ref{r1}, uncertainties which  critically affect the predictions of total ozone loss.\cite{Keutsch2025} For example, depending on the assumed reaction rate, large-scale models have shown that injecting of $5$ Mt/yr of alumina aerosol particles of $240$ nm radius could cause global ozone depletion ranging from negligible to twice the ozone loss attributed to chlorofluorocarbons in the 1980s. \cite{vattioni2023chemical}. This highlights the urgent need for a thorough analysis of SAI impact of various aerosol substrates under stratospherically relevant conditions.

The chemistry occurring on suspended particles under stratospheric conditions remains poorly understood. Solid particles such as alumina or calcite may lead to less ozone depletion than sulfuric acid aerosols\cite{keith2016stratospheric, weisenstein2015solar, WMO2022}, but available studies on particle reactivity under stratospheric conditions (i.e., temperatures below $220\,\mathrm{K}$, relative humidity below 1\%, and significant UV irradiation) often rely on simplified assumptions and/or extrapolation of experimental data at ground-level conditions. Replicating stratospheric conditions in laboratory settings is extremely challenging, making computer simulations an invaluable tool to fill these knowledge gaps. Recent MD simulations combined with surface-sensitive spectroscopy techniques (e.g., XPS) have shown that weakly hydrated surfaces (i.e., conditions that resemble those found in the stratosphere at low relative humidity) can indeed catalyze chemical reactions and enable unexpected chemical pathways compared to those observed in homogeneous  or gas-phase environments at ground-level atmospheric conditions.\cite{kong2021surface, kong2023adsorbed, faure2024formation, liu2024water} This further raises the question of whether stratospheric aerosol injection (SAI), while mitigating global warming, could also activate unintended chemical reactions.

Among the considered substrates\cite{weisenstein2015solar, pope2012stratospheric, dykema2016improved}, silica-based materials stand out due to their low cost and tunability, particularly their ability to be doped with other materials to enhance desired optical properties. Building on this idea, silica aerogels (i.e., a class of materials known for their high porosity, low density, and large surface area) doped with diamond powder yield promising results in terms of optical performance.\cite{vukajlovic2021diamond} While doped silica aerogels and related materials have been studied from an optical standpoint \cite{walker2023silica, cotterell2022accurate, vukajlovic2021diamond}, their impact on stratospheric chemistry remains largely unexplored, particularly regarding ozone depletion via chlorine activation by Reaction~\ref{r1}.

In this study, we investigate the kinetics and thermodynamics of the reaction between hydrogen chloride (\ch{HCl}) and chlorine nitrate (\ch{ClONO2}) on a dry and hydroxylated $\alpha$-quartz silica interface, a surface which serves as a proxy for silica-based aerosol substrates under low relative humidity at stratospheric conditions. We employed \emph{first-principles} MD simulations, investigating the adsorption of the reactants and products, and the kinetics of Reaction~\ref{r1}. Our results indicate a barrierless formation of \ch{Cl2} and \ch{HNO3} on the hydroxylated silica surface. Based on our MD results and available experimental data, we design and inform a kinetic model assuming a Langmuir-Hinshelwood adsorption, determining the reactive uptake coefficient for chlorine nitrate, $\gamma_{\mathrm{ClONO_2}}$, i.e., the probability that a \ch{ClONO2} collision on the silica surface results in the fomation of \ch{Cl2} by Reaction~\ref{r1}. 
$\gamma$-values are dimensionless key parameters used in defining the reaction rate coefficient, $k$, in large-scale atmospheric models, expressed as $k = \gamma \nu\, SAD / 4$. Here, $SAD$ is the aerosol surface area density (i.e., the surface area per unit volume of air) available for the reaction, which can be determined from the aerosol injection rate and particle size, while $\nu$ is the mean molecular speed of molecules (e.g., \ch{ClONO2}) colliding with the surface. Since  large atmospheric models often  calculate rates of heterogeneous reactions on atmospheric particles by simplified first-order kinetic equations~\cite{Science1997,Gao2021},  $k$ has the dimension of time$^{-1}$. 
A smaller value of $\gamma$ results in a lower heterogeneous reaction rate coefficient. 
Our kinetic model yields a negligible $\gamma_{\mathrm{ClONO_2}}$, suggesting that silica-based materials have a minimal impact on the stratospheric ozone balance.

 While this study supports silica-based materials, such as diamond-doped silica aerogels,\cite{vukajlovic2021diamond} as promising SAI technology, it also emphasizes the challenges in constraining and designing chemical kinetics at stratospheric conditions: this calls for extreme caution and further experimental validation of their chemical impact before considering their deployment in real-world SAI experiments.

\section{Methodology}

The chlorine activation via the \ch{ClONO2 + HCl} reaction has been studied here on the silica $\alpha$-quartz interface using \emph{first-principles} molecular dynamics (FPMD), i.e., the forces driving the dynamics are calculated \emph{on-the-fly} using density functional theory.
The initial $\alpha$-quartz structure was taken from previous work,\cite{D1CP04747G} and was prepared following similar protocols described in the literature.\cite{sulpizi2012silica} The system consisted of $9$ \ch{[-O-Si-O-]} layers, with two fully hydroxylated (0001) surfaces. Each surface was terminated with $16$ \ch{SiOH} silanol groups of the Q$^2$ geminal type. The z-dimension of the simulation box was enlarged to 4 nm, resulting in the formation of two vacuum/hydroxylated (0001) silica interfaces. The x-y dimensions of the simulation cell were allowed to adjust during the course of a $5$ ps semi-isotropic 1 bar constant-pressure (NpT) simulation, using a time step of $0.2$ fs. The temperature was set to 220 K, resembling the conditions found in the stratosphere. The final dimensions of the simulation box were 1.96 nm, 1.70 nm and 4.0 nm along the x, y, and z directions, respectively. The two hydroxylated surfaces were oriented normal to the z-axis, thus resembling an infinite two-dimensional $\alpha$-quartz silica slab. Finally, one \ch{HCl} and one \ch{ClONO2} molecule were placed atop one of the hydroxylated surfaces, and the system geometry was further optimized via energy minimization. A snapshot of the simulation box is shown in Supporting Information (SI) in Figure S1.

Our choice  of the $\alpha$-quartz/vacuum interface as a proxy for the silica aerosol/air interface cannot fully capture the complexity (e.g., polycrystalline structures  or amorphous substrates, such as silica aereogels) of real-world aerosols. Nevertheless, this proxy provides a reliable framework for investigating  chemical mechanisms, which are governed by atomic and molecular scale interactions, through computationally expensive quantum-mechanical calculations: such simulations offer valuable insights that can guide the design of real-world SAI.
Finally, surface hydroxylation was explicitly included, as these aerosols would likely be fabricated under ground-level conditions, where water vapor readily adsorbs and hydroxylates the surface prior to stratospheric injection.

Starting from the equilibrated silica slab surface with the two adsorbates, we performed two constant-volume and constant-temperature (NVT) molecular dynamics simulations: one at $300$ K and one at $220$ K. The simulation at $300$ K was carried out to accelerate the sampling of different structural arrangements of the two adsorbates (i.e., \ch{HCl} and \ch{ClONO2}) on the solid substrate, while the simulation at $220$ K was intended to resemble typical stratospheric temperatures.

The adsorption energies of \ch{ClONO2}, \ch{HCl}, \ch{Cl2}, and \ch{HNO3} were calculated by sampling different structural arrangements for each adsorbate on the silica surface. For each molecule, we performed 34 ps of FPMD at $220$ K following the computational procedure outlined above, starting by placing each adsorbate at the time on the silica surface. Afterwards, we extracted 30 frames from the last $30$ ps of the trajectory, selecting one frame every $1$ ps. Each snapshot was geometrically optimized, both with and without the adsorbate, via energy minimization. The adsorption energy, $E_\mathrm{ads}$, of each molecule was calculated as $E_\mathrm{ads} = E_\mathrm{compl} - (E_\mathrm{sub} + E_{X})$,
where $E_\mathrm{compl}$ is the energy of the optimized substrate with the molecule, $E_\mathrm{sub}$ is the energy of the optimized substrate without the adsorbate, and $E_X$ is the energy of the optimized molecule in the gas phase, where X corresponds to \ch{HCl}, \ch{ClONO2}, \ch{Cl2}, or \ch{HNO3}. A similar methodology for estimating adsorption energies has been adopted in the literature.\cite{D1CP04747G, doi:10.1021/jacs.4c03650, 10.1063/5.0178259}

The reaction between hydrogen chloride and chlorine nitrate was investigated by Climbing-Image Nudged Elastic Band (CI-NEB) calculations.\cite{10.1063/1.1323224, 10.1063/1.1329672} The CI-NEB method was performed using 20 replicas, where the first and last replicas corresponded to the reactants and products of Reaction~\ref{r1}, respectively. During the CI-NEB procedure, the endpoint structures, i.e., the reactants and products, were also geometrically optimized. For computational efficiency reasons, CI-NEB calculations were performed by carving snapshots of the reactants and the products from the FPMD trajectory on the silica surface, using a cutoff radius of $0.9$ nm centered on the interfacial oxygen silanol (i.e., -Si-OH) closest to \ch{ClONO2}, and terminating the structures with the \ch{Si-H} groups. Input structures for the NEB are provided in the SI, Figure S2.  The CI-NEB optimizations were then carried out in vacuum using a Martyna–Tuckerman Poisson solver.\cite{Martyna1999} The CI-NEB calculations were considered to have converged when the maximum energy variation over the last 10 replicas at any point along the energy profile did not exceed 0.5 kcal/mol. The rationale for employing CI-NEB in the study of Reaction~\ref{r1} is that, at a stratospheric temperature of 220 K and in the absence of a solvent, the entropic contribution is likely minimal, making the process primarily enthalpy-driven.

All simulations were performed using the CP2K molecular dynamics package\cite{hutter2014cp2k} The forces driving the dynamics of the system were calculated \emph{on-the-fly} using the PBE\cite{PhysRevLett.100.136406, PhysRevLett.77.3865} density functional, including Grimme’s D3 dispersion correction.\cite{doi:10.1021/acs.chemrev.5b00533} All simulations were 
carried out within an unrestricted Kohn-Sham framework, using the relaxed multiplicity scheme implemented in CP2K. Valence electrons were treated by the DZVP basis set and an energy cutoff of $500$ Ry, while core electrons were described using Goedecker–Teter–Hutter pseudopotentials.\cite{goedecker1996separable} The temperature of the system was controlled via a Nosé–Hoover thermostat with a time constant of $50$ fs.\cite{10.1063/1.463940} Time step for the FPMD simulations was set to 0.5 fs. The NEB calculations were performed both at the PBE-D3 level and using the hybrid-GGA revPBE0-D3 functional.\cite{doi:10.1021/acs.jpclett.9b01983} The revPBE0-D3 calculations were computationally accelerated by employing the auxiliary density matrix method,\cite{doi:10.1021/ct1002225} as implemented in CP2K.

\section{Results and Discussion}

\subsection{Adsorption}

\begin{figure}[ht]
    \centering
    \includegraphics[clip, trim=0cm 1cm 0cm 1cm, width=0.5\linewidth]{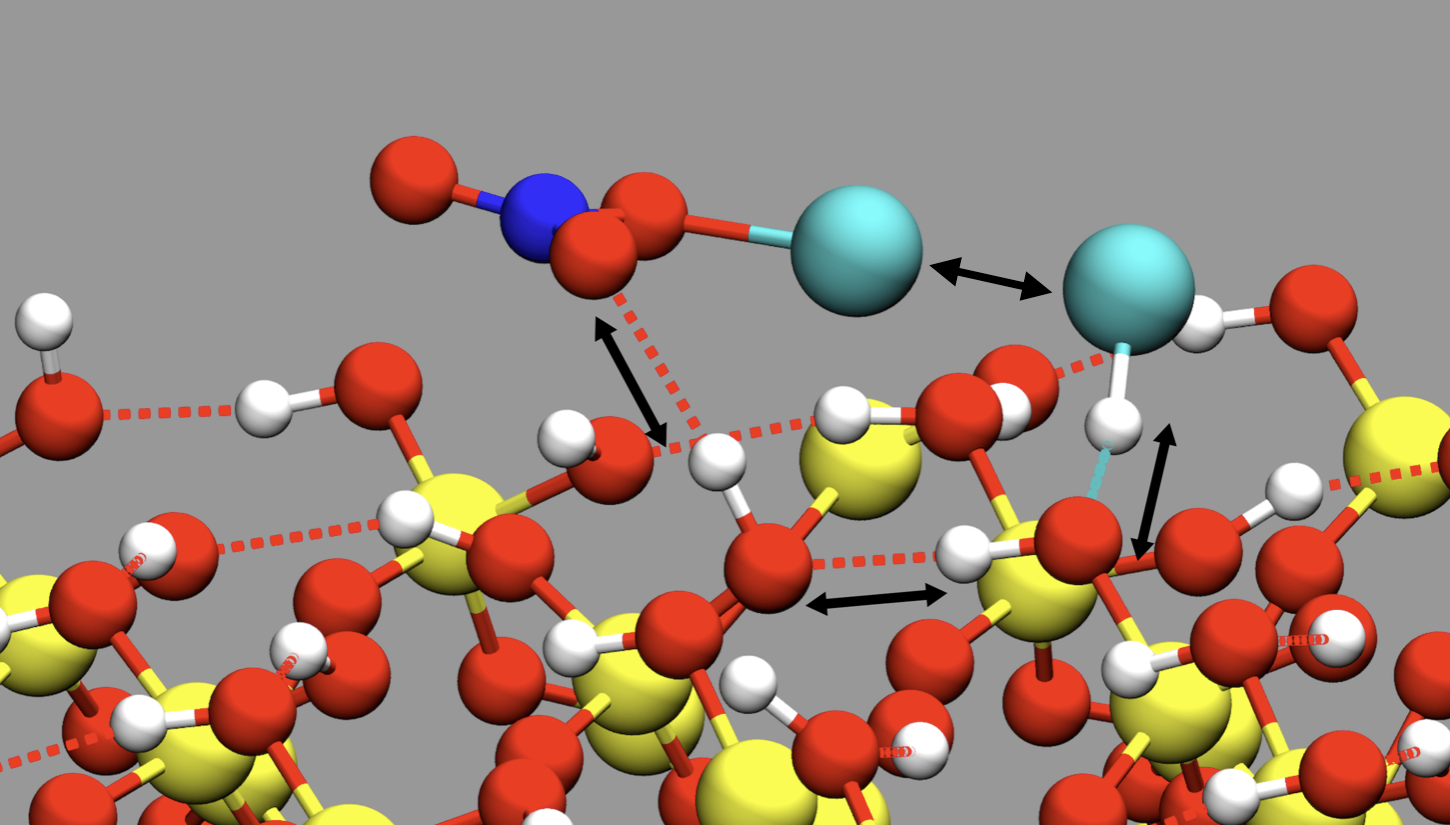}
    \caption{    
    The \ch{HCl-ClONO2} (pre)reactive structure at the hydroxylated silica surface, stabilized by hydrogen bonds (dashed red and cyan lines), observed during the FPMD simulation. A plausible mechanism for the production of nitric acid and \ch{Cl2} via proton transfer is indicated by black arrows. Atom color code: Si (yellow), Cl (cyan), O (red), H (white), N (blue).}
    \label{fig:PRE-COMPLEX}
\end{figure}

FPMD simulations at 300 K were initially performed to sample different arrangements between the reactants (i.e., \ch{HCl} and \ch{ClONO2}) on the dry and hydroxylated $\alpha$-quartz silica surface. Interestingly, the FPMD simulation at 300 K reveals the formation of an \ch{HCl-ClONO2} surface complex at around 5 ps, as shown in Figure~\ref{fig:PRE-COMPLEX}. This structural arrangement is stabilized by two hydrogen bonds with interfacial \ch{-Si-OH} silanol groups, one with the hydrogen atom of \ch{HCl} and the other with one of the oxygen atoms of chlorine nitrate. This configuration exhibits signs of a proton ring, highlighted in Figure~\ref{fig:PRE-COMPLEX} by black arrows, suggesting that this structure could serve as a plausible pre-reactive complex leading to the formation of \ch{Cl2} and nitric acid via proton transfer. At 300 K, this pre-reactive structure remains stable for at least 2 ps before the hydrogen bonds between the substrate and \ch{ClONO2} break, allowing \ch{ClONO2} to diffuse horizontally across the surface.

\begin{figure}
        \centering
        \includegraphics[ width=0.9\linewidth]{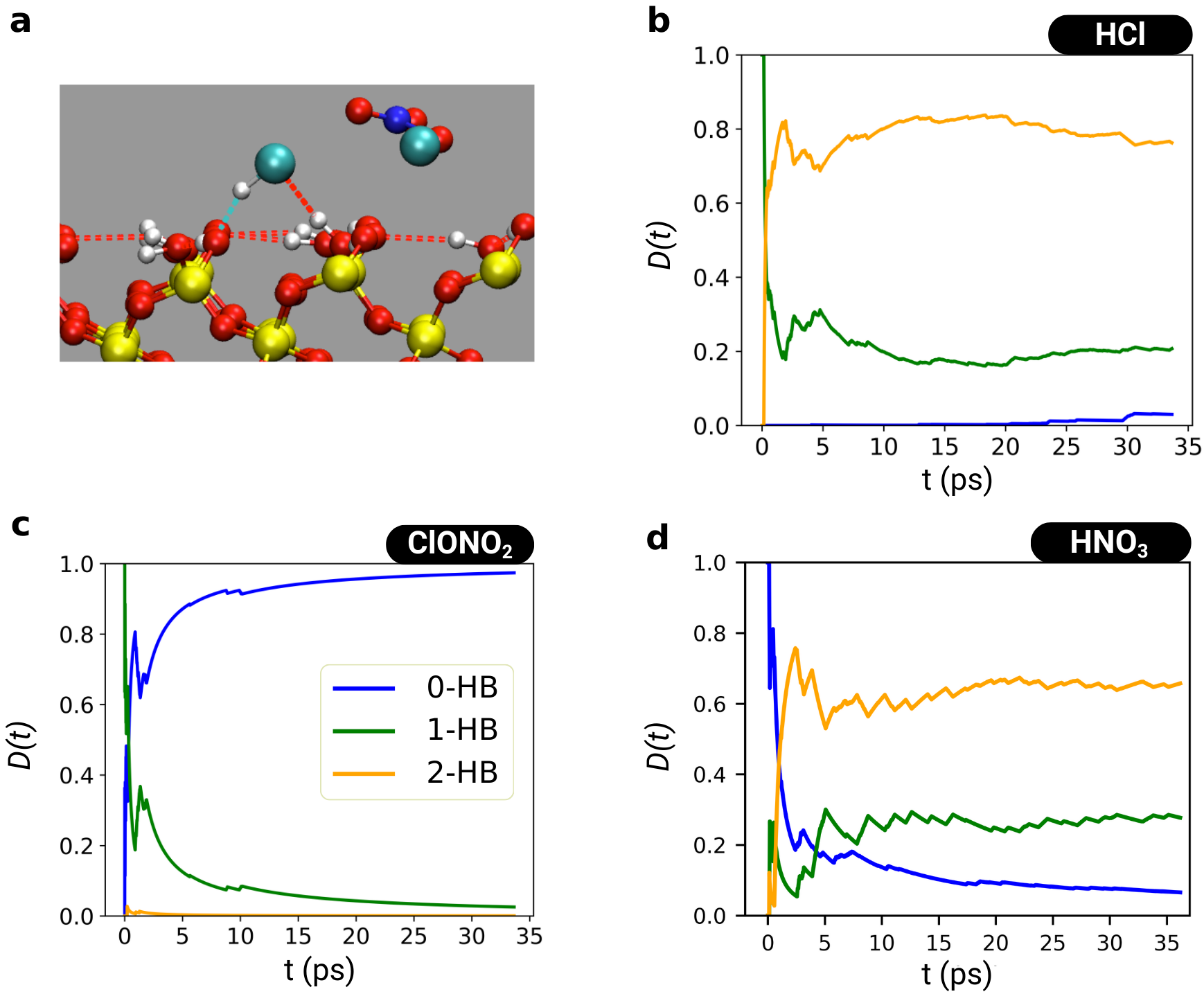}
        \caption{Panel (a): snapshot from the FPMD showing \ch{HCl} adsorbed at the surface forming two hydrogen bonds with interfacial \ch{-Si-OH} silanols. Panel (b-d), the relative cumulative (over time) hydrogen bond (HB) populations, $D(t)$, for each HB channels, i.e., 0HB (blue line), 1HB (green), and 2HB (orange). Panel (b-d) shows $D(t)$ for \ch{HCl}, \ch{ClONO2}, and \ch{HNO3}, respectivily.  Atomic color code: Si (yellow), Cl (cyan), O (red), H(white), N(blue).
        }
        \label{fig:HB}
    \end{figure}

Starting from this plausible pre-reactive structure in Figure~\ref{fig:PRE-COMPLEX}, we performed $34$ ps of FPMD at $220$ K, a temperature representative of stratospheric conditions. During this simulation, we observed a highly stable \ch{HCl} adsorbate at the interface, which remained largely immobile and did not undergo deprotonation. In contrast, \ch{ClONO2} showed greater mobility (see Figure S3). Figure~\ref{fig:HB}a illustrates the final frame at $34$ ps FPMD trajectory, showing \ch{HCl} stabilized by two hydrogen bonds. 

A more quantitative description can be provided by investigating the number of hydrogen bonds (HBs) between the adsorbates and the silica substrate. Figures~\ref{fig:HB}b-d depict the relative cumulative (over time) hydrogen bond (HB) populations, D(t), for each HB channel (i.e., 0 HB, 1 HB, 2 HB) and for \ch{HCl} and \ch{ClONO2}. $D(t)$ serves as a useful tool for identifying the most probable configurations and evaluating the statistical convergence of each HB channel. A detailed mathematical description of $D(t)$ is provided in the SI. Figure~\ref{fig:HB}b shows the $D(t)$ profile for \ch{HCl}, indicating that during approximately 80\% of the FPMD trajectory \ch{HCl} forms two hydrogen bonds with the substrate, as also visually illustrated in Figure~\ref{fig:HB}a. In the remaining 20\% of the time, \ch{HCl} forms a single hydrogen bond between its hydrogen atom and one of the oxygen of the surface silanol groups (i.e., \ch{-Si-OH}). On the other hand, \ch{ClONO2} appears to be loosely bound to the substrate, spending most of the 34 ps trajectory without forming any HB with the surface: this behavior is reflected by a larger later diffusivity of \ch{ClONO2} on the substrate compared to \ch{HCl} (see Figure S3) and, as seen later, in the different adsorption energies between the two compounds.  Figure~\ref{fig:HB}d shows the relative cumulative HB population for \ch{HNO3}, which is one of the products of reaction~\ref{r1}, suggesting as the most probable configuration for adsorbed nitric acid a structure with one or two HBs with the substrate. Although our sampling is limited due to the computational cost of expensive FPMD simulations, all HB channels (represented by orange, green, and blue lines in Figures~\ref{fig:HB}b–d) show a plateau, indicating reasonable convergence of the statistics even within relatively short duration (34 ps) of our MD trajectories.

The adsorption energies ($E_\mathrm{ads}$) of each individual reactant and product are reported in Table~{\ref{TB1}}. These values were obtained from representative snapshots sampled over four distinct FPMD trajectories (i.e., one for each of \ch{HCl}, \ch{ClONO2}, \ch{HNO3}, and \ch{Cl2}), each approximately 34 ps in length. A negative $E_\mathrm{ads}$ indicates a favorable adsorption process, and the values presented in Table~{\ref{TB1}} correlate well with the number of hydrogen bonds formed between the adsorbate and the substrate (i.e., a higher number of HBs corresponds to a more favorable adsorption). Furthermore, an estimate of the molecular residence time, $\tau$, on the substrate can be obtained using an Arrhenius-like expression: $\tau = \tau_{0} e^{-\frac{E_\mathrm{ads}}{RT}}$, where $\tau_{0}$ is typically (based on empirical observations) of the order of 1 ps.\cite{McQuarrie}  Table~{\ref{TB1}} also reports the partial pressure \ch{Cl2} recorded from space shuttle exhaust, certainly higher than those experienced under normal stratospheric conditions, while $\mathcal{F}$ is the collision flux, i.e., the number of collisions for unit of time and area. Considering that \emph{(i)} once released, \ch{Cl2} readily decomposes even in the presence of minimal sunlight over Antarctica,\cite{Molina1996} and \emph{(ii)} the residence times of \ch{HNO3} and \ch{HCl} are five and six orders of magnitude higher than those of \ch{Cl2} and \ch{ClONO2}, Table~{\ref{TB1}} suggests that the concentration of adsorbates on the silica surface will be dominated by \ch{HNO3} and \ch{HCl}, rather than \ch{Cl2} and \ch{ClONO2}.

\begin{table}[]
\caption{Adsorption energy ($E_\mathrm{ads}$), average number of HB, $\langle$HB$\rangle$, molecular residence time ($\tau$), gas-phase partial pressure at stratospheric conditions ($p_x$), and collision flux $\mathcal{F}$ obtained by kinetic theory of gas (see SI) at 220 K. [*]Values for \ch{Cl2} partial pressure from measurements from space shuttle exhaust in stratosphere.\cite{ross1997situ}}
 \label{TB1}
\begin{tabular}{lllllll}
\hline
Molecule & $E_\mathrm{ads}$  & $\langle$HB$\rangle$ & $\tau$  & p$_{x}$   & $\mathcal{F}$ \\
         & (kcal mol$^{-1}$) &    & (s)    & (Pa) & $(\mathrm{nm}^{-2}\,\mathrm{s}^{-1})$ \\
\hline \\
\ch{ClONO2}& {$-7.6 \pm 1.8$} & 0.1 & $ 4 \cdot 10^{-5}$ & $9.3 \cdot 10^{-6}$ Ref.\cite{zhang1994heterogeneous} & 0.17\\
\ch{HCl}& $-12.3 \pm 1.2$ & 1.8 & $ {2.0}$  &$ 5.0 \cdot 10^{-6}$ Ref.\cite{vattioni2023chemical}  &0.15\\
 &&&&&\\
\ch{HNO3}& $-12.9 \pm 1.7$ & 1.6 & {8.0} &  $5.1 \cdot 10^{-5}$ Ref.\cite{molina1997reaction,vattioni2023chemical} & {1.14}\\
\ch{Cl2}& $-6.6 \pm 1.2$ & 0.0 & {4.0} $\cdot 10^{-6}$ &  $6.3 \cdot 10^{-4}$ Ref.\cite{ross1997situ} $[*]$ &20.60\\
\hline
\end{tabular}
\end{table}

\subsection{Reactivity}

  FPMD simulations at 220 K and 300 K have identified a plausible pre-reactive complex for Reaction~\ref{r1}, as shown in Figure~\ref{fig:PRE-COMPLEX}, suggesting a proton transfer mechanism through interfacial silanols \ch{-Si-OH} of the substrate, which leads to hydrogen chloride deprotonation, and nitric acid and \ch{Cl2} formation. Inspired by this structure, we performed an NEB calculation to determine the reaction profile that connects reactants and products in Reaction~\ref{r1} on silica.

\begin{figure}[htbp]
    \centering
    \includegraphics[width=\textwidth]{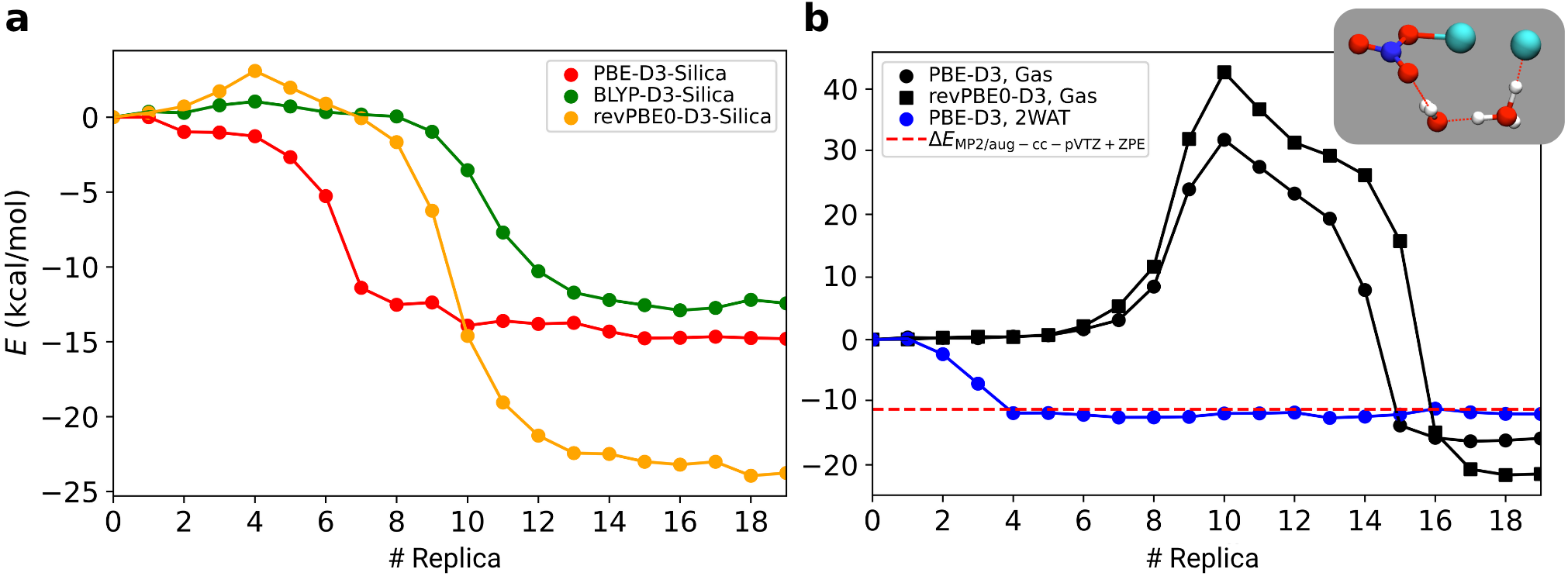}
    \caption{Panel (a): energy reaction profiles for Reaction~\ref{r1} from reactants (\ch{ClONO2} and \ch{HCl}, replica 0) to products (\ch{HNO3} and \ch{Cl2}, replica 19) on silica and at different level of density functional theory. Panel (b),  energy reaction profiles for the reaction in the gas phase and with 2 water molecules (blue dotted line). Red dashed line is $\Delta E$ for the reaction in the gas phase at MP2/aug-cc-pVTZ+ZPE level.}
    \label{fig:BARRIER}
\end{figure}

\begin{figure}[h]
    \centering
    \includegraphics[width=1\linewidth]{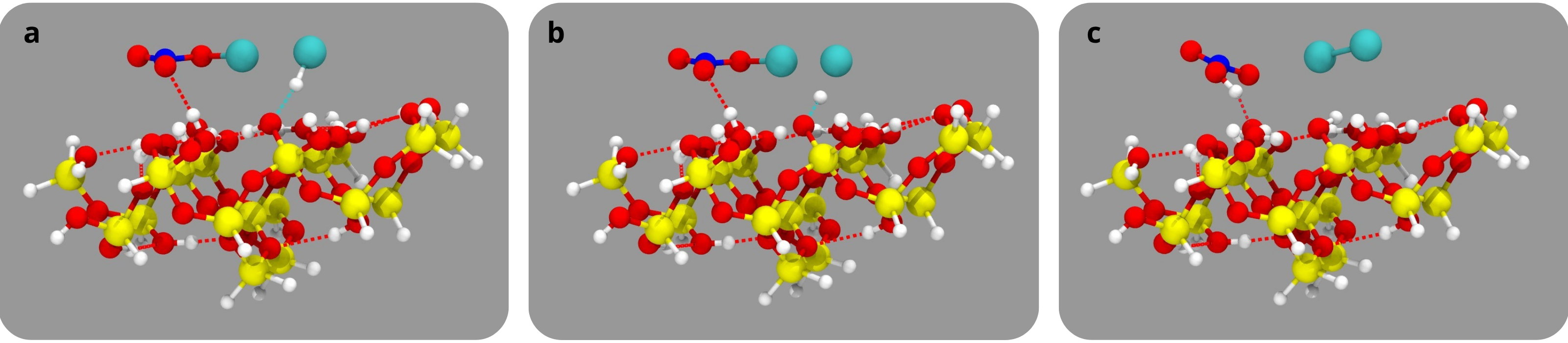}
    \caption{Reaction steps: (a) pre-reactive complex (before the barrier), (b) transition state (at the barrier), (c) post-reactive (after the barrier) from NEB calculations.}
    \label{fig:shoots}
\end{figure}

Figure~\ref{fig:BARRIER}a shows the energy profile obtained at the level of PBE-D3 (red line) using 20 replicas, which spans from reactants (replica 0) to products (replica 19). The profile reveals a thermodynamically and kinetically favorable reaction, with barrierless transition toward products and a reaction energy, $\Delta E = -15$ kcal/mol, calculated as the difference between the energies of the products and reactants. A similar conclusion can be drawn from the results obtained at the BLYP-D3 level of density functional theory (green line in Figure~\ref{fig:BARRIER}a). As a further validation, we also computed the reaction profile using the more computationally demanding hybrid revPBE0-D3 functional (orange line in Figure~\ref{fig:BARRIER}a). In the latter case, the reaction is even more exoenergetic, $\Delta E = -24$ kcal/mol, although a small transition barrier of about 3 kcal/mol is now present for the forward reaction: since the thermal energy, $\mathrm{k}_{\text{B}}T$, is approximately $0.4$ kcal/mol at $220$ K, this barrier remains relatively mild. Figure~\ref{fig:shoots} shows the different intermediates along the reaction profile. 

The picture emerging from Figure~\ref{fig:BARRIER}a indicates a forward barrierless Reaction~\ref{r1} on silica, while the reverse reaction is impeded by a significant 15 kcal/mol barrier at the PBE-D3 level and up to 25 kcal/mol at the revPBE0-D3 level. Using transition state theory, we can estimate the characteristic timescale of the reverse reaction: at 220 K this corresponds to roughly 160 seconds for the barrier calculated at the PBE-D3 level. Therefore, the backward conversion of \ch{HNO3} and \ch{Cl2} into the reactants is unlikely, particularly considering the volatility of \ch{Cl2} and its propensity to undergo photolysis.\cite{Molina1996}

Figure~\ref{fig:BARRIER}b shows reaction profiles in the gas phase, allowing a comparison with those at the hydroxylated $\alpha$-quartz interface in Figure~\ref{fig:BARRIER}a. In the gas-phase, Reaction~\ref{r1} remains thermodynamically favored, i.e. $\Delta E <0$, but the forward reaction is now kinetically hampered by a barrier of approximately 30 kcal/mol at PBE-D3 and 42 kcal/mol at revPBE0-D3 (black dotted lines and squares, respectively). As benchmark, Figure~\ref{fig:BARRIER}b  also reports the energy of reaction $\Delta E$ in the gas phase calculated at the MP2/aug-cc-pVTZ level, including Zero Point Energy (ZPE) corrections, which yields $\Delta E_{\text{MP2/aug-cc-pVTZ+ZPE}} = -11.2$ kcal/mol, value that is in quite good agreement with the $\Delta E$ at PBE-D3 level (black dotted line), and with those previously reported for Reaction~\ref{r1} in the gas phase at a similar level of density functional theory.\cite{Asada2013} 

The comparison of the reaction profiles on silica (Figure~\ref{fig:BARRIER}a) with those in the gas phase (Figure~\ref{fig:BARRIER}b) clearly underscore the substantial catalytic effect of the substrate in Reaction~\ref{r1}. This key catalytic mechanism must be attributed to the formation of an interfacial ring structure involving the reactants and the surface silanols, which facilitates proton exchange and promotes the forward reaction, as shown in Figure~\ref{fig:shoots}. To prove that, Figure~\ref{fig:BARRIER}b reports the reaction profile in the presence of two water molecules, with a starting configuration that mimics the structure of the reactant on silica (i.e., Figure~\ref{fig:shoots}a), as shown in Figure S4 and in the inset of Figure~\ref{fig:BARRIER}b: the resulting profile (blue dotted line) closely resembles that of the silica substrate (red line in Figure~\ref{fig:BARRIER}a) with a barrierless forward transition.  The findings in Figure~\ref{fig:BARRIER}b with two water molecules are not surprising and agree with previously reported ones in (very) small water clusters,\cite{McManara2000,Asada2013} showing how two water molecules are capable of lowering the forward reaction barrier from 42 kcal$/$mol in the gas phase to zero.\cite{McManara2000}
To summarize, Figure~\ref{fig:BARRIER} reinforces the proposed catalytic mechanism on silica driven by the formation of an interfacial proton ring.  

The energy reaction profiles in Figure~\ref{fig:BARRIER} have been obtained by NEB, and they do not include entropic contributions. At finite temperature, the dynamics of the system moves toward equilibrium conditions that minimize free energy, which include both enthalpic and entropic contributions. However, at stratospheric temperatures (220 K) and in the absence of solvent, which typically contributes significantly to entropy, the profiles shown in Figure~\ref{fig:BARRIER} can be reasonably interpreted as approximating free energy profiles.

\subsection{Reactive Uptake Model on Silica}

Large-scale atmospheric models have been used to estimate the impact of SAI.\cite{weisenstein2015solar,vattioni2023chemical} These models incorporate chemical subroutines that account for various heterogeneous reactive mechanisms, where one of the key parameters is the reaction probability ($\gamma$, also called as reactive uptake coefficient), that is, the probability that a gas phase molecule colliding with a surface undergoes a chemical reaction. 
Knowing \emph{(a)} the $\gamma$-values for different chemical channels, \emph{(b)} the 
available aerosol surface area density, SAD,
which can be inferred from the amount and size distribution of the injected particles, and \emph{(c)} the thermal conditions governing the mean molecular speed, $\nu$, of gas-phase molecules, it is then possible to estimate the 
heterogeneous reaction rate coefficients
as $k=\gamma SAD \nu/ 4$: these rates can be used in large scale atmospheric models to predict the impact of SAI on the atmospheric concentrations of different gas-phase species.

Vattioni et al.\cite{vattioni2023chemical} have developed a kinetic model based on available experimental data for the Reaction~\ref{r1} on alumina. Assuming a Langmuir-type adsorption and a Langmuir constant for \ch{HCl} on alumina similar to that on ice, the resulting $\gamma_{\mathrm{ClONO_2}}$ spans over two orders of magnitude, highlighting the difficulty in accurately constraining chemical mechanisms, concentrations and reaction rates under stratospheric conditions. Vattioni et al.\cite{vattioni2023chemical} also conducted a series of large-scale modeling studies, showing that the uncertainty in $\gamma$ reflects on  large variations in the predicted ozone concentrations, ranging from a negligible ozone loss to be twice to those caused by chlorofluorocarbons in the 1990s. 

Here, we assess the impact of silica-based SAI by presenting a reactive kinetic model fed by our molecular dynamics (MD) results and available literature data. To the best of our knowledge, this is the first attempt to develop a kinetic model for the heterogeneous Reaction~\ref{r1} on silica incorporating MD results to address the current lack of experimental data under stratospheric conditions. Our kinetic model is based on the following assumptions/observations:

\begin{itemize}
\item We neglect the presence of water (i.e., dry silica surface) and other trace gases (e.g., \ch{H2SO4}) in Reaction~\ref{r1}.
\item From Table~\ref{TB1}, the residence times ($\tau$) on the silica substrate before desorption are significantly longer for \ch{HCl} and \ch{HNO3} than for \ch{ClONO2} and \ch{Cl2}. Since \ch{Cl2} rapidly desorbs and is quickly photolyzed under stratospheric conditions,\cite{Molina1996} and considering the substantial backward energy barrier in Reaction~\ref{r1} (Figure~\ref{fig:BARRIER}a) the reformation of \ch{ClONO2} from \ch{Cl2} and \ch{HNO3} is highly unlikely.
\item Under stratospheric conditions, where partial pressures are low, surface coverage does not reach a full monolayer, resulting in a sparsely populated silica surface. Consequently, we assume that adsorption follows Langmuir–Hinshelwood kinetics. A Langmuir–Hinshelwood model has previously been adopted for Reaction~\ref{r1} on alumina surfaces\cite{vattioni2023chemical} and polar stratospheric clouds.\cite{ammann2013evaluated}
\item Upon adsorption on dry silica, our molecular dynamics (MD) simulations show no deprotonation of \ch{HNO3} or \ch{HCl}, and for each adsorbate the formation of (on average) of about two hydrogen bonds with adjacent silanol \ch{–Si–OH} groups. Therefore,  \ch{HCl} (or \ch{HNO3}) adsorption requires two binding sites.
\item Du et al.\cite{Du2014} experimentally estimated the Langmuir constant of \ch{HNO3} on a fused silica surface, $K_z = k^z_a/k^z_d= 107$ Torr$^{-1}$. 
\end{itemize}

From the above assumptions and insights, the backward Reaction~\ref{r1} can be neglected, and the net process proceeds according to the following steps:
\begin{align} \label{r2}
    \ch{ClONO2(g) &<=>[\ce{k^x_a}][\ce{k^x_d}]   ClONO2(ad)}
    \\ \ch{HCl(g) &<=>[\ce{k^y_a}][\ce{k^y_d}] HCl(ad)}
    \\ \ch{HCl(ad) + ClONO2(ad) &->[$\mathrm{k}^{\prime\prime}$] Cl2($\uparrow$) + HNO3(ad)}
    \\ \ch{HNO3(g) &<=>[\ce{k^{z}_a}][\ce{k^{z}_d}] HNO3(ad)}
    \label{rlast}
\end{align}

It is now possible to derive a kinetic model (see full details in the SI) based on Reactions (2)–(5) and the assumptions outlined above. This leads to a system of coupled, nonlinear differential equations for fractional coverages (i.e., $\theta$, which is the ratio between the number of occupied sites and the number of available sites on the pristine substrate) by \ch{HCl} and \ch{ClONO2} on the silica interface.  Under steady-state conditions, the system can be solved analytically using Cardano’s formula for cubic equations. Using the input data from Table 1, the kinetic model yields a value of {$\gamma_{\mathrm{ClONO_2}} = 1.0 \times 10^{-11}$}.
Silica-based materials, such as diamond-doped silica aerogels,\cite{vukajlovic2021diamond} have emerged as promising candidates for SAI due to their optical and tunable properties: our results show a negligible $\gamma_{\mathrm{ClONO_2}}$, suggesting  a minimal impact of silica-based aerosol injection on stratospheric ozone concentrations.

To identify the key molecular mechanism governing the behavior of $\gamma_{\mathrm{ClONO_2}}$, we performed a sensitivity test on our kinetic model (see Section 3.6 of the SI), examining how $\gamma_{\mathrm{ClONO_2}}$ varies with changes in the model inputs reported in Table~1. The analysis reveals that $\gamma_{\mathrm{ClONO_2}}$ is highly sensitive to the adsorption energy $E_{\mathrm{ads}}$. For example, making the \ch{ClONO2} adsorption energy more favorable by 2.4 kcal/mol (from –7.6 kcal/mol in Table~\ref{TB1} to –10 kcal/mol in Table S2) increases both $\gamma$ and the fractional coverage  of \ch{ClONO2}  by two and three orders of magnitude, respectively (see SI).
This sensitivity test likely explains why the $\gamma_{\mathrm{ClONO_2}}$ obtained here is markedly smaller than values reported in the literature (ranging from $10^{-5}$ to $10^{-1}$) for the same reaction on other substrates, such as liquid solutions, ice, doped ice, or salts.\cite{R1-1,R1-2,Crowley2010} 

For example, $\gamma_{\mathrm{ClONO_2}}$ on liquid water ranges from $10^{-5}$ to $10^{-1}$, depending on the experimental technique used,\cite{ammann2013evaluated} but \ch{ClONO2} exhibits a stronger adsorption on liquid water than on the hydroxylated $\alpha$-quartz silica surface: the adsorption energy of \ch{ClONO2} on liquid water has been reported to be between –14 and –18 kcal mol$^{-1}$,\cite{Loerting2006} compared to –7.6 kcal mol$^{-1}$ on silica (Table~\ref{TB1}).
To the best of our knowledge, only one previous study examined silica but focusing on \ch{ClONO2} hydrolysis alone and without considering the presence of HCl.\cite{R1-3}
In summary, the weak adsorption affinity of \ch{ClONO2} for silica reduces the probability of forming an \ch{HCl–ClONO2} pre-complex (see Eq. S21 in the SI), thereby resulting in a small $\gamma_{\mathrm{ClONO_2}}$.

Although our MD results enabled the development of a simplified chemical model for Reaction~\ref{r1} on the dry silica surface, it is worth recalling that obtaining $\gamma$-values for different heterogeneous chemical pathways under stratospheric conditions still remains a non-trivial task and requires a certain level of approximation and data extrapolation. Vattioni et al.\cite{vattioni2023chemical} had to use experimental data from other sources (e.g., measurements of \ch{HCl} uptake on ice) to constrain a kinetic model for Reaction~\ref{r1} on alumina and, despite this, the $\gamma_{\mathrm{ClONO_2}}$ values on alumina ranged from negligible to 0.019, high enough to contribute to stratospheric ozone depletion comparable to that observed in the late 1990s. In our kinetic model, we adopted the Langmuir constant for \ch{HNO3} from that on fused silica.\cite{Du2014} 

Furthermore, due to the inherently limited sampling of FPMD, the adsorption energies reported in Table~\ref{TB1} are subject to statistical uncertainties that can significantly affect the molecular residence time, $\tau$, given its exponential dependence on $\Delta E_\mathrm{ads}$. Although the kinetic model presented here provides a general framework applicable to other reactive uptake processes, accurate determination of adsorption energies remains crucial, as demonstrated by our sensitivity study. This requires sampling a range of structural arrangements of adsorbates, which is particularly challenging for atomistic simulations, especially at FPMD level, and accuracies on the order of $k_\mathrm{B}T$ for the adsorption energies are desirable due to the exponential scaling of desorption times. Overcoming this challenge will likely require innovative strategies to accelerate MD simulations, such as machine learning potentials or enhanced sampling techniques.\cite{sinopoli2021stability,Magrino}

Molecular dynamics simulations provide valuable insights under conditions that are difficult to probe experimentally, such as those present in the stratosphere. They can also inform kinetic models that describe reactive uptake as a set of coupled processes governed by nonlinear differential equations (e.g., Eq. S14 or Ref.\cite{Neshyba2016}), rather than relying on resistor models, which are often phenomenological and treat individual processes as uncoupled.\cite{FINLAYSONPITTS2000130, Limmer2024} 

Nevertheless, challenges remain in determining key parameters, such as reaction rates or accommodation coefficients, for large-scale atmospheric models from MD simulations, which are often unavoidably limited in sampling (as illustrated by the discussion on adsorption energies above). This call for innovative solutions and new theoretical frameworks. 

For example, Polley et al.\cite{Polley2024} recently proposed a promising approach for calculating the mass accommodation coefficient of gas molecules on dilute solutions using MD simulations alone. In their study, the dilute-phase assumption was necessary to neglect correlations between adsorbed particles, an assumption that may not hold for solid substrates where adsorbate concentrations can accumulate. Moreover, Polley et al.\cite{Polley2024} assumed a Markovian (i.e., memoryless) process, which may be questionable in the context of adsorption on solid surfaces. Bridging the gap between macroscopic observables and nanoscale simulations will require the development of new theoretical frameworks capable of integrating MD results into chemical models. This represents both an urgent and exciting area of research.

\section{Conclusions}

Given the urgency of mitigating global warming due to climate change, stratospheric aerosol injection (SAI) has been proposed as a geoengineering strategy to rapidly increase Earth's albedo. Different candidate materials have been suggested for SAI. While some exhibit favorable optical properties (i.e.,  efficiently scattering incoming shortwave solar radiation while allowing outgoing longwave radiation to escape with minimal absorption) their effects on stratospheric chemistry remain poorly understood. This also applies to  silica-based materials, such as diamond-doped silica aerogels, which show promising optical performance but whose chemical impact on stratospheric chemistry, ozone processes in particular, remains unknown.

In this study, we used a dry, hydroxylated $\alpha$-quartz silica surface as a proxy for silica-based aerosols under stratospheric conditions. We investigated the heterogeneous reaction between hydrogen chloride (\ch{HCl}) and chlorine nitrate (\ch{ClONO2}), two key reservoirs of stratospheric chlorine and nitrogen, using \emph{first-principles} molecular dynamics simulations. Our results reveal a barrierless reaction pathway leading to the formation of chlorine gas (\ch{Cl2}), a key specie contributing  to ozone depletion upon photolysis. Based on these findings and available experimental data, and by calculating the adsorption energies of \ch{HCl}, \ch{ClONO2}, \ch{HNO3}, and \ch{Cl2} on silica, we have introduced a heterogeneous kinetic model for the \ch{HCl} + \ch{ClONO2} reaction on silica. Despite the barrierless formation of \ch{Cl2}, the higher surface affinities and concentrations of \ch{HNO3} and \ch{HCl} (compared to those of \ch{ClONO2} and \ch{Cl2}) result in a negligible reactive uptake coefficient $\gamma_{\mathrm{ClONO_2}}$ upon chlorine nitrate collision on silica. These results suggest that silica-based SAI may have a limited effect on stratospheric ozone depletion.

We presented here a kinetic model informed by MD, an approach that can also complement more traditional resistor models for the determination of reactive uptakes. However,
consistent with observations from previous studies,\cite{vattioni2023chemical} this work also emphasizes the scarcity of experimental data under stratospheric conditions, which limits our ability to fully constrain kinetic models. MD simulations can fill this lack of data, but bridging the gap between molecular-scale simulations and macroscopic observables will require the development of improved theoretical frameworks and data sampling to effectively integrate MD results into kinetic modeling. Therefore, further rigorous experimental and theoretical validation is essential before silica-based SAI technologies can be confidently deployed as part of climate intervention strategies.

\section*{Acknowledgements}
 The authors thank Qatar Center for Quantum Computing, Hamad Bin Khalifa University for funding this project.  For HPC resources and services, we acknowledge the Research Computing group in Texas A$\&$M University at Qatar, founded by the Qatar Foundation for Education, Science and Community Development.
 IG would like to dedicate this work to Prof. Joseph S. Francisco on the occasion of his 70$^{th}$ birthday: a passionate scientist, inspiring teacher, esteemed colleague and dear friend.  Joe is a scientific father for many who have had the privilege of working with and learning from him. Grande Joe!

\section*{Supplementary Information}
 The Supporting Information is available free of charge in the online version of the manuscript from the publisher. The data and computational inputs supporting the findings of this study are available from the corresponding authors upon reasonable request.

\bibliographystyle{unsrt}
\bibliography{main}


\end{document}